\documentclass{article}

\usepackage[colorlinks, citecolor=blue, linkcolor=blue, backref=true]{hyperref}
\usepackage{longtable}
\usepackage{url}
\usepackage{amsthm}

\theoremstyle{definition}

\theoremstyle{remark}

\usepackage[firstinits=true, parentracker=true, doi=true, isbn=false, natbib=true]{biblatex}

\usepackage{longtable}

\usepackage[toc,page]{appendix}
\usepackage{amsmath}
\usepackage[utf8]{inputenc}
\usepackage[english]{babel}

\def\XXint#1#2#3{{\setbox0=\hbox{$#1{#2#3}{\int}$ }
\vcenter{\hbox{$#2#3$ }}\kern-.6\wd0}}

\usepackage{graphicx}
\usepackage{caption}\usepackage{subcaption}
\usepackage{url}
\usepackage{amssymb}
\usepackage{eufrak}
\usepackage{epstopdf}
\usepackage{color}

\newcommand{\vv}{\mathtt{v}}
\newcommand{\MM}{\mathtt{M}}
\newcommand{\NN}{\mathtt{N}}
\newcommand{\ctheta}{\vartheta}
\newcommand{\Nhalf}{\mathcal{N}}
\newcommand{\zphi}{\phi}
\newcommand{\Lker}{\mathtt{L}}
\newcommand{\incang}{\Theta}
\newcommand{\obsang}{\theta}
\newcommand{\xx}{x}
\newcommand{\yy}{y}
\newcommand{\uu}{u}
\newcommand{\kk}{k}
\newcommand{\oo}{\omega}
\newcommand{\varz}{z}
\newcommand{\QQ}{Q}

\let\Re\relax\DeclareMathOperator{\Re}{\text{\rm Re}}
\let\Im\relax\DeclareMathOperator{\Im}{\text{\rm Im}}

\usepackage[left]{lineno}
\setpagewiselinenumbers

\begin{document}
\title{Scattering by two staggered semi-infinite cracks on square lattice: an application of asymptotic Wiener--Hopf factorization}
\author{Gaurav Maurya\thanks{Department of Mechanical Engineering, Indian Institute of Technology Kanpur, Kanpur, U. P. 208016, India ({gmaurya@iitk.ac.in})}
\and
Basant Lal Sharma\thanks{Department of Mechanical Engineering, Indian Institute of Technology Kanpur, Kanpur, U. P. 208016, India ({bls@iitk.ac.in})}
}
\maketitle

\begin{abstract}
{Scattering of time-harmonic plane wave by two parallel semi-infinite rows, but with staggered edges, is considered on square lattice. The condition imposed on the semi-infinite rows is a discrete analogue of Neumann boundary condition. A physical interpretation assuming an out-of-plane displacement for the particles arranged in the form of a square lattice and interacting with nearest-neighbours, associates the scattering problem to lattice wave scattering due to the presence of two staggered but parallel crack tips. The discrete scattering problem is reduced to the study of a pair of Wiener--Hopf equation on an annulus in complex plane, using Fourier transforms. 
Due to the offset between the crack edges, the Wiener--Hopf kernel, a $2\times2$ matrix, is not 
amenable to factorization in a desirable form and 
an asymptotic method is adapted. 
Further, an approximation in the far field is carried out using the stationary phase method. A graphical comparison between the far-field approximation based on asymptotic Wiener--Hopf method and that obtained by a numerical solution is provided. Also included is a graphical illustration of the low frequency approximation, where it has been found that the numerical solution of the scattering problem coincides with the well known formidable solution in the continuum framework.
}
\end{abstract}

\setcounter{section}{-1}
\section{Introduction}

The solution of two dimensional Helmholtz equation with Neumann boundary condition placed on two parallel, staggered, semi-infinite edges is an intriguing problem in scattering theory \cite{abPlatesI,abPlatesII}.
A physical realization of this problem appears in the form of scattering of a time harmonic plane wave due to two hard, parallel plates placed in an acoustic medium \cite{abPlatesIII}. 
The scattering problem can be formulated as a pair of coupled Wiener--Hopf equations (due to the offset between the plates), involving a $2\times2$ matrix kernel with exponentially growing elements \cite{abPhase}. The multiplicative factorisation of such kernels in an ordinary manner \cite{noble,KreinGoh,Jones1984,MeisterSpeck} does not allow
a smooth application of Liouville's theorem, so that the Wiener--Hopf technique becomes difficult to apply \cite{Heinslim}. 
{The mathematically subtle aspects associated with such Wiener--Hopf kernels have been investigated from several viewpoints, for instance, see \cite{GohbergKaashoek}
\cite{kisil2018iterative}
\cite{mishuris2016factorization}
\cite{rogosin2015constructive}
\cite{mishuris2018regular}.}
As an approximate factorization method, in contrast to that of \cite{abPlatesI,abPlatesII},
an asymptotic method has been also proposed for such a class of problems \cite{mishu};
for instance, in the case of a small offset between the two staggered plates, the method is anticipated to yield a good approximation. 
The scattering of an anti-plane shear wave by a semi-infinite crack and that of an acoustic wave by a semi-infinite plate with hard boundary condition, are mathematically equivalent \cite{achenbook,julius} so such kernels also occur often in elastodynamics problems as well \cite{abElastic}.
Further, certain discrete analogues of the single crack diffraction problems have been also analyzed recently \cite{sK,sFK} using a square lattice model \cite{slepyan}.
In the same framework, the question of multiple scattering \cite{Meistersys1,Meistersys2,MeisterRottbrand} of a time harmonic plane wave by two staggered, semi-infinite cracks can be readily seen as an analogue of the two parallel, staggered plates problem \cite{abPlatesI, abPlatesII}. 
An analysis of such problem is the motivation for the present paper.

In this paper, the discrete scattering problem is formulated, assuming out-of-plane displacement and the presence of two semi-infinite (mode III) cracks on square lattice. The incident wave is assumed to be a time-harmonic bulk lattice wave which interacts with the two cracks and gets scattered. Using the Fourier transforms \cite{jury,sK}, a pair of coupled (due to the offset between crack tips) Wiener--Hopf equations \cite{daniele1984solution} is obtained. The Wiener--Hopf matrix kernel of the coupled equations posseses certain structure which is reminiscent of the continuum framework \cite{abPlatesI,abPlatesII}. Incidentally, we have also found that the non-zero offset case remains a difficult challenge in the square lattice framework as well. However, certain kind of incremental progress has indeed been possible since the Wiener--Hopf matrix kernel has been factorized approximately using an adaptation of an asymptotic method for the extended real line \cite{mishu} to a unit circle contour in the complex plane \cite{GMthesis}. 
Though, the existence and uniqueness of the solution has not been stated explicitly, it is anticipated that certain analogues of the statements presented by \cite{sFK} continue to hold when the imaginary part of frequency is non-zero.
The special case of zero offset is also, in itself, an interesting problem; moreover, this case admits an exact solution and has been elaborated elsewhere \cite{Bls8pair1}; an exact solution in the case of the continuum counterpart, i.e., when the scattering edges are not staggered, is well known \cite{Heins1,Heins2}. The details for the case of incidence from the waveguide formed between the cracks are omitted in the present paper, though the asymptotic factorization can be applied in the same way; only the right hand side of the Wiener--Hopf equation changes in this case. 

Besides, an illustration of the asymptotic Wiener--Hopf matrix factorization in the paper, the far-field approximation of the displacement field has been also provided assuming {\em a legitimate solution} based on the Wiener--Hopf matrix factorization. This is carried out by a standard application of the stationary phase method \cite{fokas, felsen,sK}. 
It is found that thes asymptotic factorization based far-field approximation compares with the numerical solution to some extent, specially when the offset is small relative to the spacing between the cracks. It has been found that due to an arduous task of numerical computation of certain contour integrals, there are still some difficulties with the method of asymptotic factorization of such kernels.

In the context of the well known solution of the two parallel, staggered plates problem \cite{abPlatesI, abPlatesII} in the continuum model, using the numerical solution of the discrete scattering problem, it has been found that, as the frequency approaches zero, the solution coincides with that of the the continuum model, but this is expected \cite{sConti}.

{\em Outline:} Section \ref{sqlattmodel} includes the discussion of the square lattice with two semi-infinite rows of broken bonds.
In section \ref{WHform}, with definition and application of discrete Fourier transforms,
the problem is formulated as a matrix Wiener--Hopf equation in the third section. The approximate solution of the Wiener--Hopf problem along with a discussion of the asymptotic method and factorisation of the kernel are provided thereafter in section \ref{approxWHform}. This is followed by the far-field approximation
and a discussion of the low frequency behavior in the section \ref{farfield}. 
Conclusion is presented after this and one appendix appears as well at the end of the paper.

\subsection{Notation} 
Let $\mathbb{Z}$ denote the set of integers, let $\mathbb{Z}^2$ denote $\mathbb{Z}\times\mathbb{Z}$, let $\mathbb{Z}^+$ denote the set of all non-negative integers, and let $\mathbb{Z}^-$ denote the set of all negative integers. Let $\mathbb{R}$ denote the set of real numbers, and $\mathbb{C}$ denote the set of complex numbers. The real part, $\Re{\varz}$, of a complex number $\varz\in\mathbb{C}$ is denoted by $\varz_1\in\mathbb{R}$, and its imaginary part, $\Im{\varz}$, is denoted by $\varz_2\in\mathbb{R}$ (so that $\varz=\varz_1+i\varz_2$). $|\varz|$ denotes the modulus for $\varz\in\mathbb{C}$. 
Let the letter ${H}$ stands for the Heaviside function: 
\begin{equation}
\text{${H}({\xx})=0, {\xx}\in\mathbb{Z}^-$ and ${H}({\xx})=1, {\xx\in\mathbb{Z}^+}$. }
\label{defH}
\end{equation}
The discrete Fourier transform of a sequence $\{u_m\}_{m\in\mathbb{Z}}$ is denoted by $u^F$ and defined by 
\begin{equation}
u^F(\varz)=u_+(\varz)+u_-(\varz), \quad\quad u_+(\varz)=\sum\limits_{m\in\mathbb{Z}^+}u_m\varz^{-m}, \quad\quad u_-(\varz)=\sum\limits_{m\in\mathbb{Z}^-}u_m\varz^{-m}.
\label{discFT}
\end{equation}
The symbol $\mathbb{T}$ denotes the unit circle (as a counter-clockwise contour) in the complex plane. The symbol $\varz$ is exclusively used throughout as a complex variable for the discrete Fourier transform. The square root function, $\sqrt{\cdot}$, has the usual branch cut in the complex plane running from $-\infty$ to $0$. The notation for other relevant physical and mathematical entities is described in the main text. 

\section{Square lattice model}
\label{sqlattmodel}
\begin{figure}[htb!]
\centering
\includegraphics[width=.6\linewidth]{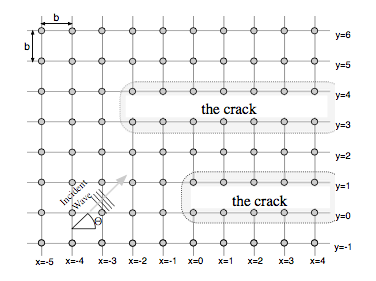}
\caption{Schematic of a pair of staggered, parallel cracks on square lattice (with $\MM=-2, \NN=3$).}
\label{cracks}
\end{figure}

Consider an infinite two-dimensional square lattice $\mathfrak{S}$ of identical particles of unit mass, which are allowed to move in the anti-plane direction. Let the displacement of a particle in $\mathfrak{S}$, indexed by its lattice coordinates $(\xx,\yy)\in\mathbb{Z}^2$, be denoted by $\uu_{\xx,\yy}\in\mathbb{C}$. Each particle in $\mathfrak{S}$ is assumed to interact with its four nearest neighbours in $\mathfrak{S}$ by linearly elastic identical (massless) bonds with a shear spring constant $1/b^2$. Assume that there are two semi-infinite cracks with broken bonds distributed between $\yy=0, \yy=1$ and $\yy=\NN, \yy=\NN+1$. Let $\Sigma_k$ denotes the set of all lattice sites that index those particles in $\mathfrak{S}$ lacking at least one nearest-neighbour bond (see Figure \ref{cracks}), that is,
\begin{equation}
\Sigma_k=\{(\xx,\yy)\in\mathbb{Z}^2:\xx\in\mathbb{Z}^+, \yy=0 \text{ or } 1\}
\cup
\{(\xx,\yy)\in\mathbb{Z}^2: \xx\in\mathbb{Z}^+, \xx\geq-\MM, \yy=\NN \text{ or }\NN+1\}.
\label{Sigmak}
\end{equation}
Suppose $\uu^i$ describes the incident lattice wave with frequency ${\boldsymbol{\omega}}$ and a lattice wave vector $(\kk_\xx,\kk_\yy)$. Specifically, it is assumed that $\uu^i$ is given by the expression
\begin{equation}
\uu^i_{\xx,\yy}=Ae^{i\kk_\xx\xx+i\kk_\yy\yy-i{\boldsymbol{\omega}} t},\quad\quad(\xx,\yy)\in\mathbb{Z}^2,
\label{uinc}
\end{equation}
where $A\in\mathbb{C}$. The explicit time dependence, $e^{-i{\boldsymbol{\omega}} t}$, is suppressed in most parts of the text.

The total displacement $\uu^t$ of an arbitrary particle in the lattice $\mathfrak{S}$ is a sum of the incident wave displacement $\uu^i$ and the scattered wave displacement $\uu$ (which includes the reflected wave)\cite{noble,sK}, that is,
\begin{equation}
\uu_{\xx,\yy}^t=\uu_{\xx,\yy}^i+\uu_{\xx,\yy},\quad\quad(\xx,\yy)\in\mathbb{Z}^2.
\label{ut}
\end{equation}
and satisfies the two dimensional discrete Helmholtz equation \cite{brill,sK}, 
\begin{equation}
\uu_{\xx+1,\yy}^t+\uu_{\xx-1,\yy}^t+\uu_{\xx,\yy+1}^t+\uu_{\xx,\yy-1}^t+(\oo^2-4)\uu_{\xx,\yy}^t=0,\quad\quad (\xx,\yy)\in\mathbb{Z}^2\setminus\Sigma_k.
\label{Heqnt}
\end{equation}

By virtue of \eqref{Heqnt} in the intact lattice, taking $\uu^t=\uu^i$, the triplet $\oo (={\boldsymbol{\omega}} b), \kk_\xx, \kk_\yy$ must satisfy the square lattice dispersion relation \cite{brill,slepyan,sFC}, namely
\begin{equation}
\oo^2=4(\sin^2\frac{1}{2}\kk_\xx+\sin^2\frac{1}{2}\kk_\yy), \quad\quad(\kk_\xx,\kk_\yy)\in[-\pi,\pi]^2.
\label{sqdispersion}
\end{equation}
In order to avoid some mathematical issues concerning Fourier transforms, a vanishingly small dissipation is assumed (see \cite{bou,noble,sC}), that is,
\begin{equation}
\oo=\oo_1+i\oo_2,\quad\quad0<\oo_2\ll1.
\label{damping}
\end{equation}
Due to the dispersion relation \eqref{sqdispersion} and the assumption \eqref{damping}, $\kk_\xx$ and $\kk_\yy$ are also complex numbers. Let $\kk$ be the lattice wave number of incident lattice wave $\uu^i$, and $\incang\in(-\pi,\pi]$, the angle of incidence (see Figure \ref{cracks}) of $\uu^i$ be defined by the relations \cite{sK}
\begin{equation}
\kk_\xx=\kk\cos\incang, \quad\quad\kk_\yy=\kk\sin\incang, \quad\quad\kk=\kk_1+i\kk_2, \quad\kk_1\geq0, \quad\quad0<\kk_2\ll1.
\label{wavenumber}
\end{equation}

Due to \eqref{ut}, \eqref{sqdispersion} and \eqref{Heqnt}, the scattered displacement field for an arbitrary particle in $\mathfrak{S}$, away from the cracks, is also governed by the discrete Helmholtz equation \eqref{Heqnt}, i.e.,
\begin{equation}
\uu_{\xx+1,\yy}+\uu_{\xx-1,\yy}+\uu_{\xx,\yy+1}+\uu_{\xx,\yy-1}+(\oo^2-4)\uu_{\xx,\yy}=0.
\label{Heqn}
\end{equation}
At the sites of broken bonds in the square lattice $\mathfrak{S}$, the equation that must be satisfied by the scattered field $\uu$ at $\yy=\NN+1$ is written as (recall \eqref{defH})
\begin{equation}
\begin{aligned}
\uu_{\xx+\MM+1,\NN+1}+\uu_{\xx+\MM-1,\NN+1}+\uu_{\xx+\MM,\NN+2}+(\oo^2-3)\uu_{\xx+\MM,\NN+1}-H(-\xx-1)\vv_{2\xx}=-H(\xx)\vv_{2\xx}^i,
\label{bcone}
\end{aligned}
\end{equation}
while at $\yy=\NN$, $\uu$ must satisfy 
\begin{equation}
\uu_{\xx+\MM+1,\NN}+\uu_{\xx+\MM-1,\NN}+\uu_{\xx+\MM,\NN-1}+(\oo^2-3)\uu_{\xx+\MM,\NN}+H(-\xx-1)\vv_{2\xx}=H(\xx)\vv_{2\xx}^i,
\label{bctwo}
\end{equation}
for all $\xx\in\mathbb{Z}$, where $H(\cdot)$ is Heaviside function \eqref{defH}. Similarly, at $\yy=1$, the scattered displacement field satisfies
\begin{equation}
\uu_{\xx+1,1}+\uu_{\xx-1,1}+\uu_{\xx,2}+(\oo^2-3)\uu_{\xx,1}-H(-\xx-1)\vv_{1\xx}=-H(\xx)\vv_{1\xx}^i,
\label{bcthree}
\end{equation}
while at $\yy=0$, the scattered displacement field satisfies
\begin{equation}
\uu_{\xx+1,0}+\uu_{\xx-1,0}+\uu_{\xx,-1}+(\oo^2-3)\uu_{\xx,0}+H(-\xx-1)\vv_{1\xx}=H(\xx)\vv_{1\xx},
\label{bcfour}
\end{equation}
for all $\xx\in\mathbb{Z}$. In \eqref{bcone} and \eqref{bcthree}
\begin{equation}
\begin{aligned}
\vv_{1\xx}&=\uu_{\xx,1}-\uu_{\xx,0}; \quad\quad\vv_{1\xx}^i=\uu_{\xx,1}^i-\uu_{\xx,0}^i;\\
\vv_{2\xx}&=\uu_{\xx+\MM,\NN+1}-\uu_{\xx+\MM,\NN}; \quad\quad\vv_{2\xx}^i=\uu_{\xx+\MM,\NN+1}^i-\uu_{\xx+\MM,\NN}^i,
\label{vs}
\end{aligned}
\end{equation}
for all $\xx\in\mathbb{Z}$, which can be approximately interpreted as scattered crack opening displacement fields and corresponding incident crack opening displacement fields.

Using \eqref{uinc}, the incident crack opening displacement fields can be written as
\begin{equation}
\vv_{1\xx}^i=e^{i\kk\xx\cos\incang}(e^{i\kk\sin\incang}-1), \quad\quad \vv_{2\xx}^i=e^{i\kk\xx\cos\incang}e^{i\kk(\NN\sin\incang+\MM \sin\incang)}(e^{i\kk\sin\incang}-1), \quad\quad \xx\in\mathbb{Z}.
\label{vinc}
\end{equation}
The equations \eqref{Heqn} through \eqref{bcfour} form the mathematical statement of the problem. The Wiener--Hopf formulation of the same is presented below.

\section{Wiener--Hopf formulation}
\label{WHform}
\subsection{Discrete Fourier transform}

Recall \eqref{wavenumber}, that is, $\kk_2>0$, so that the scattered field has a decaying nature \cite{sC}. Due to these assumptions, the discrete Fourier transform \eqref{discFT} of the field $\uu_{\xx,\yy}$ with $\xx\in\mathbb{Z}$, for any fixed $\yy\in\mathbb{Z}$, can be defined as (the details concerning the well-posedness are analogous to those presented in \cite{sK,sFK})
\begin{equation}
\uu_{\yy}^F(\varz)=\sum\limits_{\xx\in\mathbb{Z}}\uu_{\xx,\yy}\,\varz^{-\xx},\quad \varz\in\mathcal{A}_\uu,
\label{DFT}
\end{equation}
where the annulus shaped region $\mathcal{A}_\uu$ in the complex $\varz$-plane, is defined as
\begin{equation}
\mathcal{A}_\uu=\{\varz\in\mathbb{C}:R_+<|\varz|<R_-\}.
\label{annulusAU}
\end{equation}
In \eqref{annulusAU}, due to the nature of the incident wave \eqref{uinc}, we have,
\begin{equation}
R_+=e^{-\kk_2\cos\incang}, \quad\quad R_-=e^{\kk_2}.
\label{Rplusminus}
\end{equation}
The half discrete Fourier transforms \eqref{discFT} $\uu_{\yy;+}$ and  $\uu_{\yy;-}$
are analytic function of $\varz\in\mathbb{C}$ such that $|\varz|>R_+$ and $|\varz|<R_-$, respectively. 

The application of the Fourier transform \eqref{DFT} to the discrete Helmholtz equation \eqref{Heqn} gives transformed equation, which is written as
\begin{equation}
\QQ(\varz)\uu_\yy^F(\varz)-(\uu_{\yy+1}^F(\varz)+\uu_{\yy-1}^F(\varz))=0, \quad\quad \varz\in\mathcal{A}_\uu,
\label{THeqn}
\end{equation}
for all $\yy\in\mathbb{Z}$ with $\yy\neq1$, $\yy\neq0$, $\yy\neq \NN$ and $\yy\neq\NN+1$. The complex function $\QQ$ in \eqref{THeqn} is defined by 
\begin{equation}
\QQ(\varz)=4-\varz^{-1}-\varz-\oo^2,\quad\varz\in\mathbb{C}. 
\label{QHR}
\end{equation}
All the functions in \eqref{THeqn} are analytic in the annulus $\mathcal{A}_\uu$ stated by \eqref{annulusAU}. 

The equation \eqref{THeqn} is a second order difference equation \cite{levy} and its general solution is given by
\begin{equation}
\uu_\yy^F(\varz)=P(\varz)\lambda^\yy(\varz)+S(\varz)\lambda^{-\yy}(\varz),\quad\varz\in\mathcal{A},
\label{generalu}
\end{equation}
where $P$ and $S$ are arbitrary analytic function of $\varz\in\mathcal{A}$. The annulus $\mathcal{A}$ is intersection of the two annular regions in the complex $\varz$-plane (see Figure 2 of \cite{sK}), that is, using \eqref{annulusAU},
\begin{equation}
\begin{aligned}
\mathcal{A}&=\mathcal{A}_u\cap\mathcal{A}_\Lker, \\
\mathcal{A}_\Lker&=\{\varz\in\mathbb{C}:R_\Lker<|\varz|<R_\Lker^{-1}\}, \quad\quad R_\Lker=\max\{|\varz_h|,|\varz_r|\},
\label{annulus}
\end{aligned}
\end{equation}
where the pairs, $(\varz_h, \varz_h^{-1})$, and $(\varz_r, \varz_r^{-1})$, are zeros of the complex functions, $\QQ-2$ and $\QQ+2$, respectively (recall \eqref{QHR}). Following the notation and definitions of \cite{slepyan} and \cite{sK}, the complex function $\lambda$ in \eqref{generalu} is defined by 
\begin{eqnarray}
\lambda(\varz)
=\frac{r(\varz)-h(\varz)}{r(\varz)+h(\varz)}, \quad\varz\in\mathbb{C}\setminus\mathcal{B},
\label{lambda}\\
\text{where }
h(\varz)=\sqrt{\QQ(\varz)-2}, \quad\quad r(\varz)=\sqrt{\QQ(\varz)+2}, 
\label{hrlambda}
\end{eqnarray}
and $\mathcal{B}$ denotes the union of the branch cuts for $\lambda$, such that $|\lambda(\varz)|<1$, $\varz\in\mathbb{C}\setminus\mathcal{B}$. 
It should be noted that $r$ and $h$, as well as $\lambda$ are analytic in the annulus 
$\mathcal{A}$
for $\oo_2>0$ \cite{sK}. A relevant schematic of the annulus and the branch cuts was provided in the Figure 2 of \cite{sK}.

Note that the scattered displacement field decays away from the crack (recall \eqref{damping} and \eqref{wavenumber}), that is, 
\begin{equation}
\text{$\uu_\yy^F\to0$ as $\yy\to\pm\infty$,}
\end{equation}
hence, using \eqref{generalu}, the general solution to \eqref{THeqn} is expressed as 
\begin{equation}
\uu_{\yy}^F=\begin{cases}
\uu_{\NN+1}^F\lambda^{\yy-(\NN+1)},&\mbox{if } \yy\geq \NN+1, \yy\in\mathbb{Z}^+\\
A_1\lambda^{\yy}+A_2\lambda^{-\yy},&\mbox{if } 1\leq \yy\leq \NN, \yy\in\mathbb{Z},\\
\uu_{0}^F\lambda^{-\yy},&\mbox{if } \yy\in\mathbb{Z}^-\cup\{0\},\end{cases}
\label{general}
\end{equation}
where the functions $\uu_{\NN+1}^F$, $\uu_0^F$, $A_1$ and $A_2$ are unknown functions analytic in the annulus $\mathcal{A}$. 

Using \eqref{general}, $\uu_1^F$ and $\uu_\NN^F$ can be written in terms of $A_1$ and $A_2$, that is,
$\uu_{1}^F=A_1\lambda+A_2\lambda^{-1}; 
\uu_{\NN}^F=A_1\lambda^{\NN}+A_2\lambda^{-\NN},$
where $\uu_{1}^F$ and $\uu_{\NN}^F$ are unknown functions, analytic in the annulus $\mathcal{A}$.
The unknowns $A_1$ and $A_2$ can be thus replaced by $\uu_1^F$ and $\uu_{\NN}^F$ after solving these two equations,
that is, 
\begin{equation}
A_1=\frac{(\uu_{\NN}^F\lambda^{-1}-\uu_1^F\lambda^{-\NN})}{\lambda^{\NN-1}-\lambda^{-\NN+1}}; \quad\quad 
A_2=\frac{(\uu_1^F\lambda^{\NN}-\uu_{\NN}^F\lambda)}{(\lambda^{\NN-1}-\lambda^{-\NN+1})}.
\label{At}
\end{equation}
Substitution of these expressions in the general solution \eqref{general}, $\uu_{\NN-1}^F$ and $\uu_{2}^F$ can be also written in terms of $\uu_\NN^F$ and $\uu_{1}^F$, that is,
\begin{equation}
\uu_{\NN-1}^F=A_1\lambda^{\NN-1}+A_2\lambda^{1-\NN}=\frac{\uu_\NN^F(\lambda^{\NN-2}-\lambda^{2-\NN})+\uu_{1}^F(\lambda-\lambda^{-1})}{\lambda^{\NN-1}-\lambda^{-\NN+1}},
\label{UNm}
\end{equation}
and 
\begin{equation}
\uu_{2}^F=A_1\lambda^{2}+A_2\lambda^{-2}=\frac{\uu_\NN^F(\lambda-\lambda^{-1})+\uu_{1}^F(\lambda^{\NN-2}-\lambda^{-\NN+2})}{\lambda^{\NN-1}-\lambda^{-\NN+1}},
\label{Utwo}
\end{equation}
respectively. 

Using the manipulations and basic expressions stated above, the matrix Wiener--Hopf equation associated with the presence of two semi-infinite rows of broken bonds is derived in the subsequent portion of this section. 

\subsection{Matrix Wiener--Hopf equation}
The scattering problem for two staggered cracks on square lattice is formulated as coupled Wiener--Hopf equations \cite{noble,KreinGoh}. 
Let
\begin{equation}
\varz_{\text{P}}=e^{i\kk\cos\incang}\in\mathbb{C}; \quad\quad
\delta_{D+}(\varz)=\sum\limits_{\xx\in\mathbb{Z}^+}\varz^{-\xx}, \varz\in\mathbb{C}, \text{with } |\varz|>1.
\label{zp}
\end{equation}
By its definition, the function $\delta_{D+}(\varz\varz_{\text{P}}^{-1})$ is analytic in the region exterior to the circle of radius $|\varz_{\text{P}}|$, in the complex $\varz$-plane, such that it has a pole at $\varz=\varz_{\text{P}}$ 
in the complex plane. Using the definition \eqref{zp}, the discrete Fourier transforms of \eqref{vinc} are given by
\begin{equation}
\vv_{1+}^i=A(e^{i\kk\sin\incang}-1)\delta_{D+}(\varz\varz_{\text{P}}^{-1}); \quad
\vv_{2+}^i=e^{i\kk(\MM \cos\incang+\NN\sin\incang)}(e^{i\kk\sin\incang}-1)\delta_{D+}(\varz\varz_{\text{P}}^{-1}),
\label{Tincident}
\end{equation}
for all $\varz\in\mathbb{C}, \text{with } |\varz|>\max\{R_+,R_\Lker\}$. Thus, $\vv_{1+}^i$ and $\vv_{2+}^i$ are analytic on the annulus $\mathcal{A}$ and outside it.

Using the 
($\xx$-shift) properties of the discrete Fourier transform \cite{esaber}, also recall \eqref{QHR}, 
the equations \eqref{bcone} through \eqref{bcfour} (suppressing the argument $\varz$ of functions) lead to
\begin{subequations}
\begin{eqnarray}
\uu_{\NN+1}^F(\QQ-1)-\uu_{\NN+2}^F+\varz^{-\MM}\vv_{2-}&=&\varz^{-\MM}\vv_{2+}^i,
\label{Tbcone}\\
\uu_{\NN}^F(\QQ-1)-\uu_{\NN-1}^F-\varz^{-\MM}\vv_{2-}&=&-\varz^{-\MM}\vv_{2+}^i,
\label{Tbctwo}\\
\uu_{1}^F(\QQ-1)-\uu_{2}^F+\vv_{1-}&=&\vv_{1+}^i,
\label{Tbcthree}\\
\uu_{0}^F(\QQ-1)-\uu_{-1}^F-\vv_{1-}&=&-\vv_{1+}^i,
\label{Tbcfour}
\end{eqnarray}
\end{subequations}
respectively, for all $\varz\in\mathcal{A}$.

The expressions \eqref{UNm} and \eqref{Utwo} is substituted in \eqref{Tbctwo} and \eqref{Tbcthree}, to get 
\begin{equation}
\uu_\NN^F(\QQ-1)-\frac{\uu_\NN^F(\lambda^{\NN-2}-\lambda^{2-\NN})+\uu_{1}^F(\lambda-\lambda^{-1})}{\lambda^{\NN-1}-\lambda^{-\NN+1}}-\varz^{-\MM}\vv_{2-}=-\varz^{-\MM}\vv_{2+}^i,
\label{UN}
\end{equation}
and
\begin{equation}
\uu_{1}^F(\QQ-1)-\frac{\uu_\NN^F(\lambda-\lambda^{-1})+\uu_{1}^F(\lambda^{\NN-2}-\lambda^{-\NN+2})}{\lambda^{\NN-1}-\lambda^{-\NN+1}}+\vv_{1-}=\vv_{1+}^i,
\label{Uone}
\end{equation}
for all $\varz\in\mathcal{A}$. The equations \eqref{UN} and \eqref{Uone} can be solved for $\uu_{\NN}^F$ and $\uu_1^F$, in terms of $\mathtt{v}_{2-}$, $\mathtt{v}_{2+}^i$, $\mathtt{v}_{1-}$ and $\mathtt{v}_{1+}^i$. Substituting these expressions in the general solution \eqref{general} and using the general solution in \eqref{Tbcone} and \eqref{Tbcfour} along with the definitions \eqref{vs}, the coupled Wiener--Hopf equations are found to be
\begin{equation}
\mathbf{v}_-+\mathbf{K}\mathbf{v}_++\mathbf{f}^i=0, \quad\varz\in\mathcal{A},
\label{MatrixWH}
\end{equation}
where 
\begin{subequations}
\begin{eqnarray}
\mathbf{v}_-&=&\begin{bmatrix}
\vv_{1-}\\\vv_{2-}
\end{bmatrix}; \quad\quad
\mathbf{v}_+=\begin{bmatrix}
\vv_{1+}\\\vv_{2+}\end{bmatrix};\\
\mathbf{K}&=&\Lker\begin{bmatrix}
1&\varz^{-\MM}\lambda^\NN\\
\varz^{\MM}\lambda^\NN&1
\end{bmatrix}; \quad\quad\Lker(\varz)=\frac{h(\varz)}{r(\varz)}, 
\label{defKL}\\
\mathbf{f}^i&=&-\frac{1-e^{i\kk\sin\incang}}{1+\lambda}\begin{bmatrix}
2\lambda&\varz^{-\MM}\lambda^\NN(\lambda-1)\\\varz^{\MM}\lambda^\NN(\lambda-1)&2\lambda\end{bmatrix}\delta_{D+}(\varz\varz_{\text{P}}^{-1})\begin{bmatrix}1\\e^{i\kk(\NN\sin\incang+\MM \cos\incang)}\end{bmatrix}.
\label{matrix}
\end{eqnarray}
\label{WHeqn}
\end{subequations}
Recall \eqref{hrlambda} that $h=\sqrt{\QQ-2}; r=\sqrt{\QQ+2}$.
In \eqref{matrix}, the matrix function $\mathbf{K}(\varz)$ is the $2\times2$ Wiener--Hopf kernel which needs to be multiplicatively factorised with appropriate behaviour of the factors in their respective regions of analyticity for the application of the Wiener--Hopf technique \cite{noble}. The issue of a desirable factorisation of $\mathbf{K}(\varz)$ and the formal solution of the Wiener--Hopf equation \eqref{MatrixWH} are discussed in the next section.

\section{Approximate solution of the Wiener--Hopf equation}
\label{approxWHform}

A fundamental step in Wiener--Hopf technique is the factorisation of the kernel such that the factors have certain behaviour in their respective regions of analyticity in the complex plane \cite{noble}. In case of the kernel $\mathbf{K}$, the ordinary methods of factorisation of a matrix, result into factors with behaviour inappropriate for the application of the Liouville's theorem. At the moment, there is no constructive method for this task, in order to tackle this issue, an asymptotic method is adapted from a recently published work 
\cite{mishu,bottcher}. For this purpose, the scheme for an extended real line \cite{mishu} needs to be mapped to a unit circle contour in the complex plane \cite{GMthesis}; some details have been omitted as they are direct analogues of the former. 

\subsection{Asymptotic method of factorisation}

Let $\mathbb{T}_{+}$ and $\mathbb{T}_-$ be, respectively, the exterior and the interior of the unit circle $\mathbb{T}$ in the complex plane. Let $\mathcal{C}(\mathbb{T})$ be the set of all continuous functions on $\mathbb{T}$. Let $\mathcal{C}_{+}(\mathbb{T})$ and $\mathcal{C}_-(\mathbb{T})$ stand for the subset of those functions in $\mathcal{C}(\mathbb{T})$ that admit continuous extensions onto $\mathbb{T}\cup\mathbb{T}_+$ and $\mathbb{T}\cup\mathbb{T}_-$, which are analytic in $\mathbb{T}_+$ and $\mathbb{T}_-$, respectively.
The matrix Wiener--Hopf kernel $\mathbf{K}$ falls in the class $\mathfrak{G}_2$ \cite{mishu}. Let
\begin{equation}
\mathbf{G}_\MM(\varz)=\begin{bmatrix}
1& \varz^{-\MM}\lambda^{\NN}(\varz)\\
\varz^{\MM}\lambda^{\NN}(\varz)&1
\end{bmatrix},
\label{GM}
\end{equation}
Then, \eqref{GM} can be written in the form of 
$\mathbf{G}_\MM=\mathbf{R}_\MM\mathbf{ F} \mathbf{R}_\MM^{-1}$ with
\begin{equation}
\mathbf{R}_\MM(\varz)=\begin{bmatrix}
\varz^{-(\frac{\MM}{2})}&0\\0&\varz^{(\frac{\MM}{2})}
\end{bmatrix},
\label{RM}
\end{equation}
so that when $\MM=0$, $\mathbf{R}_\MM(\varz)|_{\MM=0}=\mathbf{I}$, and 
\begin{equation}
\mathbf{F}(\varz)=\begin{bmatrix}
1&\lambda^\NN(\varz)\\
\lambda^\NN(\varz)&1
\end{bmatrix}
\label{Fz}.
\end{equation}
It is stated without proof that $\mathbf{R}_\MM$ is analytic, bounded and locally H{\"o}lder-continuous \cite{evans,gilbarg} 
on $\mathbb{T}$.
The complex function $\lambda$ has branch cuts in the complex plane with zeros of $h$ and $r$ as the branch points (see \eqref{lambda}). But these branch points do not lie on the unit circle $\mathbb{T}$, as can be seen in Figure 
2 of \cite{sK}. Since, the branch cuts are selected such that $|\lambda(\varz)|<1, \forall \varz\in\mathbb{T}$,
\begin{equation}
\text{$\det\mathbf{F}(\varz)=1-\lambda^{2\NN}$, is non-zero on $\mathbb{T}$ }
\end{equation}
and 
\begin{equation}
\text{ind}\det \mathbf{F}(\varz)= 0.
\end{equation}
Furthermore, the eigenvalues $1\pm \lambda^{\NN}$ and hence, the determinant is positive, so that, $\mathbf{F}(\varz)$ is positive-definite. 
We have the canonical factorization \cite{kaashoek,mishu}
\begin{equation}
\mathbf{F}(\varz)=\mathbf{F}_-(\varz)\mathbf{F}_+(\varz),
\end{equation}
as $\mathbf{F}(\varz)$ is invertible for all $\varz\in\mathbb{T}$. Symbolically, the matrix function $\mathbf{F}(\varz)$ can be written as 
\begin{eqnarray}
\mathbf{F}(\varz)&=&\mathbf{P}\begin{bmatrix}
G_1(\varz)&0\\
0&G_2(\varz)
\end{bmatrix}\mathbf{P},
\label{factorF}\\
\text{where }
G_1(\varz)&=&1+\lambda^\NN(\varz),\quad\quad G_2(\varz)=1-\lambda^\NN(\varz),\quad\quad \mathbf{P}=\frac{1}{\sqrt{2}}\begin{bmatrix}
1&1\\1&-1
\end{bmatrix}.
\label{matrixP}
\end{eqnarray}
Using the Cauchy projectors (see \cite{sK,sC,gakhov}), 
\begin{equation}
G_{j\pm}(\varz)=\exp(\pm\frac{1}{2\pi i}\oint_{\mathbb{T}}\frac{\log(G_{j}(\alpha))}{(\varz-\alpha)}\,d\alpha),\quad\quad \varz\in\mathbb{C}, j=1, 2,
\label{Gtwopm}
\end{equation}
where the integrals 
can be calculated numerically,
however, the two functions $G_1$ and $G_2$ can also simplified further and then factorised. The factorisation, for the case {\em when $\NN$ is even}, is detailed in the Appendix \ref{app_elementary}.
We obtain the canonical factorisation of the matrix function $\mathbf{F}(\varz)$ \eqref{factorF} with the factors given by
\begin{equation}
\begin{aligned}
\mathbf{F}_-(\varz)=\frac{1}{\sqrt{2}}\begin{bmatrix}
G_{1-}(\varz)&G_{2-}(\varz)\\
G_{1-}(\varz)&-G_{2-}(\varz)
\end{bmatrix}; \\
\mathbf{F}_+(\varz)=\frac{1}{\sqrt{2}}\begin{bmatrix}
G_{1+}(\varz)&G_{1+}(\varz)\\
G_{2+}(\varz)&-G_{2+}(\varz)
\end{bmatrix}.
\label{Fminusplusfact}
\end{aligned}
\end{equation}
Let the determinants of matrix functions $\mathbf{F}_+(\varz)$ and $\mathbf{F}_-(\varz)$ be denoted by $\Delta_+(\varz)$ and $\Delta_-(\varz)$, respectively. They are expressed as
\begin{equation}
\Delta_-(\varz)=-G_{1-}(\varz)G_{2-}(\varz); \quad\quad\Delta_+(\varz)=-G_{1+}(\varz)G_{2+}(\varz).
\label{det}
\end{equation}
The inverse matrices of the factors $\mathbf{F}_-$ and $\mathbf{F}_+$, respectively, are 
\begin{equation}
\begin{aligned}
\mathbf{F}_-^{-1}(\varz)=\frac{1}{\sqrt{2}}\begin{bmatrix}
G_{1-}^{-1}(\varz)&G_{1-}^{-1}(\varz)\\
G_{2-}^{-1}(\varz)&-G_{2-}^{-1}(\varz)
\end{bmatrix}; \\
\mathbf{F}_+^{-1}(\varz)=\frac{1}{\sqrt{2}}\begin{bmatrix}
G_{1+}^{-1}(\varz)&G_{2+}^{-1}(\varz)\\
G_{1+}^{-1}(\varz)&-G_{2+}^{-1}(\varz)
\end{bmatrix}.
\label{Fminusplusinv}
\end{aligned}
\end{equation}
Above factorization for $\MM=0$ is essentially the matrix counterpart of the analysis presented in \cite{Bls8pair1} where symmetry has been used to reduce the matrix Wiener--Hopf problem to a scalar Wiener--Hopf.

Using \eqref{GM} 
and \eqref{Fminusplusinv}, 
the complex matrix function $\mathbf{G}_{1,\MM}$ defined by 
\begin{equation}
\mathbf{G}_{1,\MM}(\varz)=\mathbf{F}_-^{-1}(\varz)\mathbf{G}_\MM(\varz)\mathbf{F}_+^{-1}(\varz)
\end{equation}
can be re-written as 
\begin{equation}
\mathbf{G}_{1,\MM}=\begin{bmatrix}
\dfrac{2+\lambda^{\NN}(\varz)(\varz^{\MM}+\varz^{-\MM})}{2G_1(\varz)}&\dfrac{\lambda^\NN(\varz)(\varz^\MM-\varz^{-\MM})}{2G_{1-}(\varz)G_{2+}(\varz)}\\
\dfrac{\lambda^\NN(\varz)(\varz^{-\MM}-\varz^{\MM})}{2G_{1+}(\varz)G_{2-}(\varz)}&\dfrac{2-\lambda^{\NN}(\varz)(\varz^\MM+\varz^{-\MM})}{2G_2(\varz)}
\end{bmatrix}.
\label{GoneM}
\end{equation}
At this point, with 
\begin{equation}
\varz=e^{-i\xi}
\end{equation}
in \eqref{GoneM}, above is modified to
\begin{equation}
\mathbf{G}_{1,\MM}=\begin{bmatrix}
\dfrac{1+\lambda^{\NN}(\xi)\cos{\xi \MM}}{G_1(\xi)}&-i\dfrac{G_{1+}(\xi)G_{2-}(\xi)\lambda^\NN(\xi)\sin{\xi \MM}}{G_{1}(\xi)G_{2}(\xi)}\\
i\dfrac{G_{1-}(\xi)G_{2+}(\xi)\lambda^\NN(\xi)\sin{\xi \MM}}{G_{1}(\xi)G_{2}(\xi)}&\dfrac{1-\lambda^{\NN}(\xi)\cos{\xi \MM}}{G_2(\xi)}
\end{bmatrix}.
\label{GMone}
\end{equation}
The functions on the diagonal of the matrix \eqref{GMone} can be rewritten as
\begin{equation}
\begin{aligned}
\frac{1+\lambda^{\NN}(\xi)\cos{\xi M}}{G_1(\xi)}=1-2\frac{\lambda^\NN(\xi)\sin^2(\xi \MM/2)}{G_1(\xi)}, \\
\frac{1-\lambda^{\NN}(\xi)\cos{\xi \MM}}{G_2(\xi)}=1+2\frac{\lambda^\NN(\xi)\sin^2(\xi \MM/2)}{G_2(\xi)}.
\end{aligned}
\end{equation}
Finally, \eqref{GoneM} can be expressed such that
\begin{equation}
\mathbf{G}_{1,\MM}(\xi)=\mathbf{I}+\epsilon\mathbf{\tilde{N}}_\MM(\xi),
\label{GoneMN}
\end{equation}
\text{where }
\begin{equation}
\epsilon\mathbf{\tilde{N}}_\MM(\xi) = \lambda^\NN(\xi)\sin(\frac{1}{2}\xi \MM)\begin{bmatrix}
-\dfrac{2\sin(\frac{1}{2}\xi \MM)}{G_1(\xi)}&-i\dfrac{2G_{1+}(\xi)G_{2-}(\xi)\cos(\frac{1}{2}\xi \MM)}{G_{1}(\xi)G_{2}(\xi)}\\i\dfrac{2G_{1-}(\xi)G_{2+}(\xi)\cos{(\frac{1}{2}\xi \MM)}}{G_{1}(\xi)G_{2}(\xi)}&\dfrac{2\sin(\frac{1}{2}\xi \MM)}{G_2(\xi)}
\end{bmatrix}.
\label{factorGone}
\end{equation}
With 
\begin{equation}
\epsilon\equiv\epsilon(\xi)= \lambda^\NN(\xi)\sin(\xi \MM/2),
\end{equation}
due to the branch selection for $\lambda(\varz)$. 
Reverting back to the $\varz$ based formulation, the first order approximation of the $\mathbf{G}_{1,\MM}$ can be written as 
\begin{equation}
\mathbf{G}_{1,\MM}(\varz)=\mathbf{I}+\mathbf{N}_\MM(\varz)=(\mathbf{I}+\mathbf{N}_{1\MM-}(\varz))(\mathbf{I}+\mathbf{N}_{1\MM+}(\varz)),
\label{firstOGfact}
\end{equation}
where
\begin{equation}
\mathbf{N}_\MM(\varz)=\begin{bmatrix}
-\dfrac{\lambda^\NN(\varz)(2-\varz^\MM-\varz^{-\MM})}{2G_1(\varz)}&\dfrac{G_{1+}(\varz)G_{2-}(\varz)\lambda^\NN(\varz)(\varz^\MM-\varz^{-\MM})}{2G_1(\varz)G_2(\varz)}\\
-\dfrac{G_{1-}(\varz)G_{2+}(\varz)\lambda^\NN(\varz)(\varz^\MM-\varz^{-\MM})}{2G_1(\varz)G_2(\varz)}&\dfrac{\lambda^\NN(\varz)(2-\varz^\MM-\varz^{-\MM})}{2G_2(\varz)}
\end{bmatrix}.
\label{NMz}
\end{equation}
In the additive expansion,
\begin{equation}
\mathbf{N}_{1\MM+}(\varz)+\mathbf{N}_{1\MM-}(\varz)=\mathbf{N}_\MM(\varz),
\label{addfactN}
\end{equation}
the factors, $\mathbf{N}_{1\MM+}(\varz)$ and $\mathbf{N}_{1\MM-}(\varz)$, are given by 
$\mathbf{N}_{1{\MM}\pm}(\varz)=\pm\frac{1}{2\pi i}\oint_{\mathbb{T}}\frac{\mathbf{N}_{\MM}({\alpha})}{\varz-\alpha}\,d\alpha, \varz\in\mathbb{T}$, respectively. 
Since, all the singularities of the functions in $\mathbf{N}_\MM(\varz)$ are either inside or outside the unit circle $\mathbb{T}$, the integrals 
can be performed numerically. 
Hence, the first order approximate multiplicative factorisation of the matrix function $\mathbf{G}_\MM(\varz)$ is given by
\begin{equation}
\mathbf{G}_\MM(\varz)=\mathbf{F}_-(\varz)(\mathbf{I}+\mathbf{N}_{1\MM-}(\varz))(\mathbf{I}+\mathbf{N}_{1\MM+}(\varz))\mathbf{F}_+(\varz).
\label{firstGfact}
\end{equation}

\subsection{Approximate solution of the scattering problem through Wiener--Hopf technique}

Once the matrix kernel is appropriately factorised, the Wiener--Hopf equations \eqref{MatrixWH} can be solved using the Wiener--Hopf technique. Using the factorisation \eqref{firstGfact}, the factorised kernel is written as 
\begin{equation}
\mathbf{K}(\varz)=\Lker(\varz)\mathbf{G}_\MM(\varz)=\mathbf{K}_{-}(\varz) \mathbf{K}_{+}(\varz),\quad\varz\in\mathcal{A}
\label{kKfact}
\end{equation}
where the first order factors are given by
\begin{equation}
\mathbf{K}_{-}(\varz)=\Lker_-(\varz)\mathbf{F}_-(\varz)(\mathbf{I}+\mathbf{N}_{1\MM-}(\varz)), \varz\in\mathbb{C}, \text{with} |\varz|<\min\{R_-,R_L^{-1}\},
\label{kKminus}
\end{equation}
and 
\begin{equation}
\mathbf{K}_{+}(\varz)=\Lker_+(\varz)(\mathbf{I}+\mathbf{N}_{1\MM+}(\varz))\mathbf{F}_+(\varz), \varz\in\mathbb{C}, \text{with} |\varz|>\max\{R_+,R_L\},
\label{kKplus}
\end{equation}
with $\Lker$ defined by \eqref{defKL}${}_3$.
The function $\Lker$ does not vanish on a unit circle in $\mathbb{T}$ in the complex plane and also, the $\text{ind }\Lker=0$ on $\mathbb{T}$, which are the sufficient conditions for factorisation of the function on the unit circle \cite{sFK}, that is, 
\begin{equation}
\Lker_\pm(\varz)=\exp(\pm\frac{1}{2\pi i}\int\limits_{\mathbb{T}}\frac{\log\Lker(\alpha)}{\varz-\alpha}\,d\alpha), \quad\quad\varz\in\mathbb{C}, \text{with} |\varz|\lessgtr R_\Lker^{\pm1}.
\label{Lpm}
\end{equation}
The factorisation of the function $\Lker(\varz)$ is carried out in \cite{sK} and the explicit expressions for the factors are given by
\begin{equation}
\Lker_+(\varz)=\Lker_-(\varz^{-1})=\mathcal{C}_\Lker\sqrt{\frac{1-\varz_h\varz^{-1}}{1-\varz_r\varz^{-1}}}, \quad\quad\varz\in\mathbb{C}, \text{such that} |\varz|>R_L,
\label{explicitL}
\end{equation}
with
\begin{equation}
\mathcal{C}_L=(\varz_r/\varz_h)^{\frac{1}{4}}\in\mathbb{C}.
\end{equation}
Due to the factorisation \eqref{kKfact}, the coupled Wiener--Hopf equation \eqref{MatrixWH} is modified as 
\begin{equation}
\mathbf{K}_{-}^{-1}(\varz)\mathbf{v}_-(\varz)+\mathbf{K}_{+}(\varz)\mathbf{v}_+(\varz)+\mathbf{K}_{-}^{-1}(\varz)\mathbf{f}^i(\varz)=0,\quad\varz\in\mathcal{A}.
\label{factKk}
\end{equation}
The kernel $\mathbf{K}(\varz)$ can be written as 
\begin{equation}
\mathbf{K}(\varz)=\mathbf{I}+({\lambda(\varz)+1})^{-1}\begin{bmatrix}
-2{\lambda(\varz)}&-\varz^{-\MM}\lambda^\NN(\varz)({\lambda(\varz)-1})\\
-\varz^{\MM}\lambda^\NN(\varz)({\lambda(\varz)-1})&-2{\lambda(\varz)}
\end{bmatrix}
\end{equation}
The equation \eqref{factKk} can be rewritten as 
\begin{equation}
\mathbf{K}_{-}^{-1}(\varz)\mathbf{v}_-(\varz)+\mathbf{K}_{+}(\varz)\mathbf{v}_+(\varz)+\mathbf{K}_{-}^{-1}(\varz)(\mathbf{K}(\varz)-\mathbf{I})\mathbf{v}^i(\varz)=0,\quad\varz\in\mathcal{A},
\label{kkfact}
\end{equation}
or by using \eqref{kKfact},
\begin{equation}
\mathbf{K}_{-}^{-1}(\varz)\mathbf{v}_-(\varz)+\mathbf{K}_{+}(\varz)\mathbf{v}_+(\varz)+(\mathbf{K}_{+}(\varz)-\mathbf{K}_{-}^{-1}(\varz))\mathbf{v}^i(\varz)=0,\quad\varz\in\mathcal{A},
\label{kkfact}
\end{equation}
where
\begin{equation}
\mathbf{v}^i(\varz)=-(1-e^{i\kk\sin\incang})\delta_{D+}(\varz\varz_{\text{P}}^{-1})\begin{bmatrix}1\\e^{i\kk(\NN\sin\incang+\MM\cos\incang)}\end{bmatrix},
\end{equation}
for all $\varz\in\mathbb{C}$ with $|\varz|>\max\{R_+,R_\Lker\}$. The equation \eqref{kkfact} can be rearranged and rewritten as
\begin{equation}
\mathbf{K}_{-}^{-1}(\varz)\mathbf{v}_-(\varz)+\mathbf{K}_{+}(\varz)\mathbf{v}_+(\varz)=\mathbf{C}(\varz),\quad\varz\in\mathcal{A},
\label{KKfact}
\end{equation}
where 
\begin{equation}
\mathbf{C}(\varz)=(\mathbf{K}_{-}^{-1}(\varz)-\mathbf{K}_{+}(\varz))\mathbf{v}^i(\varz),\quad\varz\in\mathcal{A}.
\label{Cmat}
\end{equation}

\begin{figure}[htb!]
\centering
\includegraphics[width=.5\linewidth]{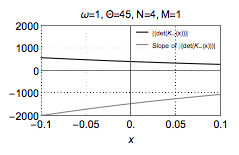}
\caption{Determinant of $	\mathbf{K}_{k-}(\varz)$ as $\varz\to0$, with $\varz=\varz(x)=x+0.005i$, $x$ on the horizontal axis. The black curve shows the determinant while the grey plot shows the slope of the determinant. The parameters chosen for plotting purpose are shown in the plot label.}
\label{DeterminantKminus}
\end{figure}

\begin{figure}[htb!]
\centering
\includegraphics[width=\linewidth]{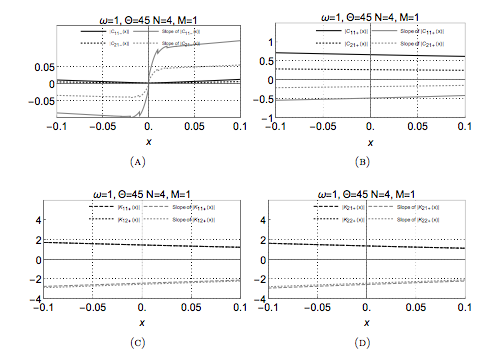}
\caption{The behaviour of the elements of (A) $\mathbf{C}_{-}(\varz)$, (B) $\mathbf{C}_{+}(1/\varz)$ and (C, D) $\mathbf{K}_{+}(1/\varz)$ as $\varz\to0$, with $\varz=\varz(x)=x+0.005i$, $x$ on the horizontal axis.}
\label{Cminus}
\end{figure}

The function $\delta_{D+}(\varz\varz_{\text{P}}^{-1})$ has a simple pole at $\varz=\varz_{\text{P}}$, which lies outside the annulus of analyticity $\mathcal{A}$ but inside the unit circle in complex plane. The vector function $\mathbf{C}$ can be additively factorised as 
\begin{equation}
\mathbf{C}(\varz)=\mathbf{C}_+(\varz)+\mathbf{C}_-(\varz);
\end{equation}
the factors are given by
\begin{equation}
\begin{aligned}
\mathbf{C}_+(\varz)=(-\mathbf{K}_{+}(\varz)+\mathbf{K}_{-}^{-1}(\varz_{\text{P}}))\mathbf{v}^i(\varz);\\
\mathbf{C}_-(\varz)=(\mathbf{K}_{-}^{-1}(\varz)-\mathbf{K}_{-}^{-1}(\varz_{\text{P}}))\mathbf{v}^i(\varz),
\label{factC}
\end{aligned}
\end{equation}
where $\mathbf{C}_+$ and $\mathbf{C}_+$ are analytic at $\varz\in\mathbb{C}$ with $|\varz|>\max\{R_+,R_\Lker\}$, $|\varz|<\min\{R_-,R_\Lker^{-1}\}$, respectively. Using the factorisation \eqref{factC} in \eqref{KKfact}, we have
\begin{equation}
\mathbf{K}_{-}^{-1}(\varz)\mathbf{v}_-(\varz)-\mathbf{C}_-(\varz)=-\mathbf{K}_{+}(\varz)\mathbf{v}_+(\varz)+\mathbf{C}_+(\varz)\quad\equiv\mathbf{J}(\varz).
\label{completefactKK}
\end{equation}
Since the two sides are analytic continuation of each other in their respective planes of analyticity, the function $\mathbf{J}(\varz)$ is an entire function on the complex plane. 

Examining \eqref{Goneplusminus}, \eqref{Gtwoplusminus}, and \eqref{NMz}, 
numerically, it can be found that the behaviour of the functions in \eqref{completefactKK}, is suitable for application of the Liouville's theorem. 
The numerically obtained plots of the function have been shown in the Figure \ref{DeterminantKminus} and Figure \ref{Cminus}. 

By inspection, it can be seen that the functions on the left of equation \eqref{completefactKK} tend to zero as $\varz$ tends to zero whereas the right hand side is a constant as $\varz$ approaches infinity. Hence, the functions have appropriate behaviour for application of the Louville's theorem and that, $\mathbf{J}(\varz)=\boldsymbol{0}$ and therefore,
\begin{equation}
\mathbf{v}_-(\varz)=\mathbf{K}_{-}(\varz)\mathbf{C}_-(\varz), \quad\quad\varz\in\mathbb{C} \text{with} |\varz|<\min\{R_-,R_\Lker^{-1}\},
\label{solmWHone}
\end{equation}
and
\begin{equation}
\mathbf{v}_+(\varz)=\mathbf{K}_{+}^{-1}(\varz)\mathbf{C}_+(\varz), \quad\quad\varz\in\mathbb{C} \text{with} |\varz|>\max\{R_+,R_\Lker\}.
\label{solmWHtwo}
\end{equation}
Then, from the relation $\mathbf{v}^F(\varz)=\mathbf{v}_-(\varz)+\mathbf{v}_+(\varz)$, $\varz\in\mathcal{A}$, by adding \eqref{solmWHone} and \eqref{solmWHtwo}, $\mathbf{v}^F(\varz)$ has the following expression
\begin{equation}
\mathbf{v}^F(\varz)=(\mathbf{K}_{+}^{-1}(\varz)-\mathbf{K}_{-}(\varz))\mathbf{K}_{-}^{-1}(\varz_{\text{P}})\mathbf{v}^i(\varz).
\label{solWHk}
\end{equation}
The functions given by \eqref{solmWHone} and \eqref{solmWHtwo} form the solution of the Wiener--Hopf equation \eqref{MatrixWH} in the form of a discrete Fourier transform. The wave field in the physical lattice can be obtained by inverting the Fourier transform, so that the approximate solution of the scattering problem can be written in integral form. For this purpose, the equations \eqref{Tbcone} and \eqref{Tbcfour},
\begin{align}
\uu_{\NN+1}^F(\lambda^{-1}-1)+\varz^{-\MM}\vv_{2-}&=\varz^{-\MM}\vv_{2+}^i,
\label{Tbconemodone}\\
\uu_{0}^F(\lambda^{-1}-1)-\vv_{1-}&=-\vv_{1+}^i,
\label{Tbcfourmodeone}
\end{align}
are written in the matrix form, that is,
\begin{subequations}
\begin{eqnarray}
\uu_{\NN+1}^F(\lambda^{-1}-1)+\varz^{-\MM}
\mathbf{a}^\top\begin{bmatrix}\vv_{1-}\\\vv_{2-}\end{bmatrix}&=&\varz^{-\MM}\vv_{2+}^i,
\label{Tbconemodtwo}\\
\uu_{0}^F(\lambda^{-1}-1)-
\mathbf{b}^\top\begin{bmatrix}\vv_{1-}\\\vv_{2-}\end{bmatrix}&=&-\vv_{1+}^i,
\label{Tbcfourmodetwo}\\
\text{where }
&&\mathbf{a}^\top=\begin{bmatrix} 0&1\end{bmatrix}\text{ and }\mathbf{b}^\top=\begin{bmatrix} 1&0\end{bmatrix}.
\label{defab}
\end{eqnarray}
\end{subequations}
Using the solution of the Wiener--Hopf equation, \eqref{solmWHone}, the equations \eqref{Tbconemodtwo} and \eqref{Tbcfourmodetwo} can be written in terms of the factor of kernel $\mathbf{K}$, that is,
\begin{eqnarray}
\uu_{\NN+1}^F&=&-\varz^{-\MM}
\frac{(1-e^{i\kk\sin\incang})\delta_{D+}(\varz\varz_{\text{P}}^{-1})}{(\lambda^{-1}-1)}\mathbf{a}^\top\mathbf{K}_{-}(\varz)\mathbf{K}_{-}^{-1}(\varz_{\text{P}})\mathbf{f},
\label{TBCONEMOD}\\
\text{and }
\uu_0^F&=&
\frac{(1-e^{i\kk\sin\incang})\delta_{D+}(\varz\varz_{\text{P}}^{-1})}{(\lambda^{-1}-1)}\mathbf{b}^\top\mathbf{K}_{-}(\varz)\mathbf{K}_{-}^{-1}(\varz_{\text{P}})\mathbf{f},
\label{TBCFOURMOD}
\end{eqnarray}
where 
\begin{equation}
\mathbf{f}=\begin{bmatrix}1\\e^{i\kk(\NN\sin\incang+\MM\cos\incang)}\end{bmatrix}.
\label{deff}
\end{equation}
In above equations following definitions have been used
\begin{equation}
\begin{aligned}
\vv_{1+}^i&=(e^{i\kk\sin\incang}-1)\delta_{D+}(\varz\varz_{\text{P}}^{-1}), \\
\vv_{2+}^i&=e^{i\kk(\MM\cos\incang+\NN\sin\incang)}(e^{i\kk\sin\incang}-1)\delta_{D+}(\varz\varz_{\text{P}}^{-1}),
\label{defin}
\end{aligned} 
\end{equation}
and
\begin{equation}
\mathbf{v}^i=-(1-e^{i\kk\sin\incang})\delta_{D+}(\varz\varz_{\text{P}}^{-1})\begin{bmatrix}1\\e^{i\kk(\NN\sin\incang+\MM\cos\incang)}\end{bmatrix}.
\end{equation}
The expressions \eqref{TBCONEMOD} and \eqref{TBCFOURMOD} along with \eqref{general} give
\begin{equation}
\uu_{\yy}^F=(-\varz^{-\MM}\frac{(1-e^{i\kk\sin\incang})\delta_{D+}(\varz\varz_{\text{P}}^{-1})}{(\lambda^{-1}-1)}\mathbf{a}^\top\mathbf{K}_{-}(\varz)\mathbf{K}_{-}^{-1}(\varz_{\text{P}})\mathbf{f})\lambda^{\yy-(\NN+1)}, \quad\quad\yy\geq \NN+1,
\label{solutionone}
\end{equation} 
and
\begin{equation}
\uu_{\yy}^F=(\frac{(1-e^{i\kk\sin\incang})\delta_{D+}(\varz\varz_{\text{P}}^{-1})}{(\lambda^{-1}-1)}\mathbf{b}^\top\mathbf{K}_{-}(\varz)\mathbf{K}_{-}^{-1}(\varz_{\text{P}})\mathbf{f})\lambda^{-\yy}, \quad\quad\yy\leq0.
\label{solutiontwo}
\end{equation}
Expressions \eqref{solutionone} and \eqref{solutiontwo} can be further simplified as $\delta_{D+}(\varz\varz_{\text{P}}^{-1})=\frac{\varz}{\varz-\varz_{\text{P}}}$ in the annulus. Hence,
\begin{equation}
\uu_{\yy}^F=(-\varz^{-\MM}\frac{(1-e^{i\kk\sin\incang})}{(\lambda^{-1}-1)}\mathbf{a}^\top\mathbf{K}_{-}(\varz)\mathbf{K}_{-}^{-1}(\varz_{\text{P}})\mathbf{f})\frac{\lambda^{\yy-(\NN+1)}\varz}{\varz-\varz_{\text{P}}}, \quad\quad\yy\geq \NN+1,
\label{solutionone}
\end{equation} 
and
\begin{equation}
\uu_{\yy}^F=(\frac{(1-e^{i\kk\sin\incang})}{(\lambda^{-1}-1)}\mathbf{b}^\top\mathbf{K}_{-}(\varz)\mathbf{K}_{-}^{-1}(\varz_{\text{P}})\mathbf{f})\frac{\lambda^{-\yy}\varz}{\varz-\varz_{\text{P}}}, \quad\quad\yy\leq0.
\label{solutiontwo}
\end{equation}
The discrete Fourier transform can be inverted and the scattered displacement field can be obtained. For $\yy\geq \NN+1$, 
\begin{equation}
\uu_{\xx,\yy}=-\frac{C_0}{2\pi i}\oint_{C_\varz}\mathbf{a}^\top(\mathbf{K}_{-}(\varz)\mathbf{K}_{-}^{-1}(\varz_{\text{P}}^{-1})\mathbf{f})\frac{\lambda^{\yy-\NN}(\varz)\varz^{\xx-\MM}}{(\varz-\varz_{\text{P}}^{-1})(1-\lambda(\varz))}\,d\varz,
\label{uxynone}
\end{equation}
while for $\yy\leq 0$, 
\begin{equation}
\uu_{\xx,\yy}=\frac{C_0}{2\pi i}\oint_{C_\varz}\mathbf{b}^\top(\mathbf{K}_{-}(\varz)\mathbf{K}_{-}^{-1}(\varz_{\text{P}}^{-1})\mathbf{f})\frac{\lambda^{-\yy+1}(\varz)\varz^{\xx}}{(\varz-\varz_{\text{P}}^{-1})(1-\lambda(\varz))}\,d\varz,
\label{uxyzone}
\end{equation}
where 
\begin{equation}
C_0=1-e^{i\kk\sin\incang}
\label{defC0}
\end{equation}
and $C_\varz$ is a circular contour lying inside the annulus of analyticity in the $\varz$-complex plane. \eqref{uxynone} and \eqref{uxyzone} provide the complete solution of the scattering problem in the integral form.

\section{Approximate description of the far-field behavior
for the Two Crack Problem}
\label{farfield}
This section closely follows the analysis presented in \cite{sK}. The approximation of far-field can be obtained using the stationary phase method \cite{fokas, felsen}. 
Utilising the mapping 
\begin{equation}
\varz=e^{-i\xi}, \quad\quad\varz_{\text{P}}=e^{-i\xi_{\text{P}}},
\end{equation}
we can write the expressions \eqref{uxynone} and \eqref{uxyzone} as 
\begin{equation}
\uu_{\xx,\yy}=-\frac{C_0}{2\pi i}\int\limits_{C_\xi}\mathbf{a}^\top(\mathbf{K}_{-}(e^{-i\xi})\mathbf{K}_{-}^{-1}(\xi_{\text{P}})\mathbf{f})\frac{\lambda^{\yy-\NN}(e^{-i\xi})e^{-i\xi(\xx-\MM)}}{(e^{-i\xi}-e^{-i\xi_{\text{P}}})(1-\lambda(e^{-i\xi}))}(-ie^{-i\xi})\,d\xi,
\label{uxyntwo}
\end{equation}
\begin{equation}
\uu_{\xx,\yy}=\frac{C_0}{2\pi i}\int\limits_{C_\xi}\mathbf{b}^\top(\mathbf{K}_{-}(e^{-i\xi})\mathbf{K}_{-}^{-1}(\xi_{\text{P}})\mathbf{f})\frac{\lambda^{-\yy+1}(e^{-i\xi})e^{-i\xi\xx}}{(e^{-i\xi}-e^{-i\xi_{\text{P}}})(1-\lambda(e^{-i\xi}))}(-ie^{-i\xi})\,d\xi,
\label{uxyztwo}
\end{equation}
respectively, for $\yy\geq \NN+1$ and $\yy\leq 0$, where $C_\xi$ is a 
contour of finite length (with $\xi$ traversed from $-\pi$ to $\pi$) lying inside the strip of analyticity in the $\xi$-complex plane 
\begin{equation}
\mathcal{S}=\{\xi\in\mathbb{C}:\xi\in[-\pi,\pi],-\kk_2\cos\incang<\xi_2<\kk_2\}.
\label{StripS}
\end{equation}
The mapping between $(\xx,\yy)$ and $(R,\obsang)$ with 
\begin{equation}
\xx=R\cos\obsang
\text{ and }\yy=R\sin\obsang,
\label{polarcoord}
\end{equation}
$\obsang\in(0,\pi)$ for $\yy\geq\NN+1$ and $\obsang\in(\pi,2\pi)$ for $\yy\leq0$ can be used for further analysis.
With \eqref{defK} and \eqref{defG}, the relation
\begin{equation}
\lambda(e^{-i\xi})=e^{i\eta(e^{-i\xi})}\equiv e^{i\eta(\xi)},
\end{equation}
and the polar coordinates, 
\eqref{uxyntwo} and \eqref{uxyztwo}, respectively,
can be written as
\begin{equation}
\uu_{\xx,\yy}=-\frac{C_0}{2\pi}\int\limits_{C_\xi}\mathcal{K}(e^{-i\xi})\frac{e^{-i\eta\NN}e^{iR{\zphi}_2(\xi)}e^{i\xi\MM}}{(e^{i(\xi-\xi_{\text{P}})}-1)(1-e^{i\eta})}\,d\xi, \quad\quad(\yy\geq \NN+1)
\label{uxynseven}
\end{equation}
\begin{equation}
\uu_{\xx,\yy}=-\frac{C_0}{2\pi }\int\limits_{C_\xi}\mathcal{G}(e^{-i\xi})\frac{e^{i\eta}e^{iR{\zphi}_1(\xi)}}{(1-e^{i(\xi-\xi_{\text{P}})})(1-e^{i\eta})}\,d\xi \quad\quad(\yy\leq 0).
\label{uxyzseven}
\end{equation}
where the phase functions ${\zphi}_2$ and ${\zphi}_1$ are given by
\begin{equation}
{\zphi}_2(\xi)=\eta\sin\obsang-\xi\cos\obsang, \quad\quad{\zphi}_1(\xi)=-\eta\sin\obsang-\xi\cos\obsang, \xi\in\mathcal{S}.
\label{phiinstrip}
\end{equation}

The functions in \eqref{phiinstrip} possess saddle points \cite{fokas,felsen,mittra,harris,sK} at $\xi=\xi_{s2}$ and $\xi=\xi_{s1}$, respectively on $C_\xi$, given by
\begin{align}
{\zphi}_1'(\xi_{s1})&=-\eta'(\xi_{s1})\sin\obsang-\cos\obsang=0, \quad\quad{\zphi}_1''(\xi_{s1})=-\eta''(\xi_{s1})\sin\obsang\neq0,
\label{xisone}\\
{\zphi}_2'(\xi_{s2})&=\eta'(\xi_{s2})\sin\obsang-\cos\obsang=0, \quad\quad{\zphi}_2''(\xi_{s2})=\eta''(\xi_{s2})\sin\obsang\neq0.
\label{xistwo}
\end{align}
Following \cite{sK}, $\eta(\xi)=\cos^{-1}(\varpi-\cos\xi)$, where $\varpi=2-\frac{1}{2}\oo^2$, $\xi\in C_\xi$. Using these, the equations \eqref{xisone} and \eqref{xistwo}, become
\begin{equation}
\frac{\sin{\xi_{s1}}}{\sqrt{1-(\varpi-\cos\xi)^2}}=\cot\obsang, \quad\quad\obsang\in(\pi,2\pi),
\label{xioneeq}
\end{equation}
and
\begin{equation}
\frac{\sin{\xi_{s2}}}{\sqrt{1-(\varpi-\cos\xi)^2}}=-\cot\obsang, \quad\quad\obsang\in(0,\pi).
\label{xitwoeq}
\end{equation}
For simplicity, let us consider only the case $\oo\in(0,2)$ (the details in case $\oo\in(2,2\sqrt{2})$ follow alterations similar to those presented by \cite{sK} for a single crack).
Assuming $\obsang\in(0,\pi/2)$, from \eqref{xitwoeq}, it follows that $\xi_{s2}\in(-\pi,0)$, and $\xi_{s2}\in(0,\pi)$, when $\obsang\in(\pi/2,\pi)$. Hence,
$\xi_{s2}=\begin{cases}
-\Xi(\obsang)&\mbox{if} \obsang \in(0,\pi/2),\\
+\Xi(\obsang)&\mbox{if} \obsang\in(\pi/2,0),
\end{cases}$
where
$\Xi(\obsang)=\cos^{-1}\frac{1}{2}(\varpi+\tau(\obsang)), \obsang\in[0,2\pi],$
and
$\tau(\obsang)=\sec2\obsang(\varpi\pm\sqrt{\varpi^2\sin^22\obsang+4\cos^22\obsang}),$
Now assuming $\obsang\in(\pi,3\pi/2)$, from \eqref{xioneeq}, it follows that $\xi_{s1}\in(0,\pi)$, and $\xi_{s1}\in(-\pi,0)$, when $\obsang\in(3\pi/2,2\pi)$. Hence, 
$\xi_{s1}=\begin{cases}
+\Xi(\obsang)&\mbox{if} \obsang \in(\pi,3\pi/2),\\
-\Xi(\obsang)&\mbox{if} \obsang\in(3\pi/2,2\pi).
\end{cases}$
As $\oo\in(0,2)$, it can be seen that the pair $\varz_h^\pm$ lies close to the unit circle $\mathbb{T}$ in $\mathbb{C}$, that is, when $\oo$ is close to 0, $\varz_h$ is close to 1 and when $\oo$ is close to 2, $\varz_h$ is close to -1. Hence, $\varz_h^\pm$ can be written as $e^{\pm i\xi_h}$ with $\xi_h\in(0,\pi)$. By definition, $\varz_{\text{P}}=e^{-i\kk\cos\incang}$. This makes 
\begin{equation}
\xi_{\text{P}}=\kk\cos\incang.
\end{equation}
Now using the mapping $\xi=\xi_h\cos\alpha$, the integrals \eqref{uxynseven} and \eqref{uxyzseven} can be transformed into
\begin{equation}
\uu_{\xx,\yy}=\frac{C_0\xi_h}{2\pi}\int\limits_{C_\alpha}\mathcal{K}(e^{-i\xi_h\cos\alpha})\frac{e^{-i\eta(\xi_h\cos\alpha)\NN}e^{iR{\zphi}_2(\xi_h \cos\alpha)}e^{i\xi_h\MM \cos\alpha}}{(e^{i(\xi_h\cos\alpha-\xi_{\text{P}})}-1)(1-e^{i\eta(\xi_h\cos\alpha)})}\sin\alpha\,d\alpha,\quad\quad(\yy\geq \NN+1)
\label{uxynalpha}
\end{equation}
and 
\begin{equation}
\uu_{\xx,\yy}=\frac{C_0\xi_h}{2\pi }\int\limits_{C_\alpha}\mathcal{G}(e^{-i\xi_h\cos\alpha})\frac{e^{i\eta(\xi_h\cos\alpha)}e^{iR{\zphi}_1(\xi_h\cos\alpha)}}{(1-e^{i(\xi_h\cos\alpha-\xi_{\text{P}})})(1-e^{i\eta(\xi_h\cos\alpha)})}\sin\alpha\,d\alpha,\quad\quad(\yy\leq 0)
\label{uxyzalpha}
\end{equation}
where the endpoints of the contour $C_\alpha$ are given by $\xi_h\cos\alpha_i=-\pi$ and $\xi_h\cos\alpha_f=+\pi$, with $\alpha_i=\pi-ia_i$ and $\alpha_f=0+ia_f$, where $a_i\geq0$ and $a_f\geq0$ and it can be seen that $a_i=a_f$ (calculated numerically). The contour $C_\alpha$ starts at $\alpha_i$ and ends at $\alpha_f$ and can be deformed on steepest descent contour (passing from $\alpha_s$, the saddle point) with or without the contribution of the the pole $\alpha_{\text{P}}$ if $\alpha_s\lessgtr\alpha_{\text{P}}$.

\begin{figure}[htb!]
\centering
\includegraphics[width=\linewidth]{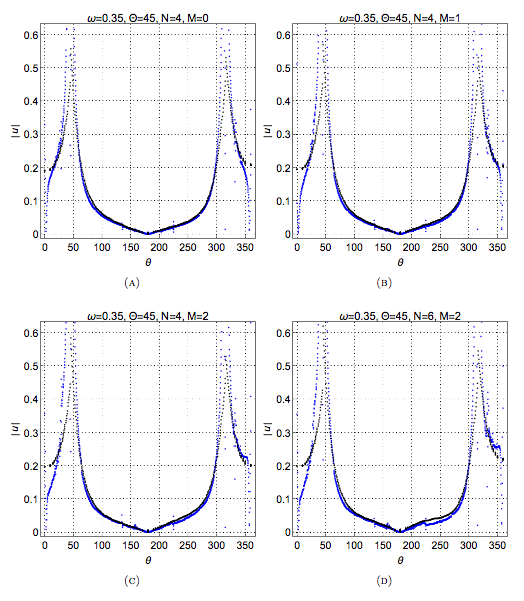}
 \caption{Modulus of the diffracted field $|\uu|$ against the observation angle $\theta$ \eqref{polarcoord} for angle of incidence $\Theta=45^\circ$, $\NN=4$ and (A) $\MM=0,$ (B) $\MM=1$, (C) $\MM=2$ and (D) $\NN=6$ and $\MM=2$. The numerical results are shown in black while the semi-analytical results are shown in blue. The other parameters used are $\oo=0.35$, $N_{grid}=448$, $N_{pml}=270$. The radius of the discrete circle is 70 according to the relation \eqref{polarcoord}.}
 \label{twocracks}
\end{figure}

After deforming the contour $C_\alpha$ to $C_{\alpha_s}$, it is found that modulo the contribution of pole
\begin{equation}
\uu_{\xx,\yy}\sim\uu_{\xx,\yy}|_s,
\label{approxS}
\end{equation}
as $\xi_hR\to\infty$, where,
for $\yy\geq\NN+1$, 
\begin{equation}
\uu_{\xx,\yy}|_s\sim-\sqrt{\frac{i}{R|{\zphi}_2''(\alpha_s)|}}\frac{C_0}{\sqrt{2\pi}}\mathcal{K}(\varz_s)\frac{e^{-i\eta(\xi_s)\NN}e^{iR{\zphi}_{2}(\xi_s)}e^{i\xi_s\MM }}{(e^{i(\xi_s-\xi_{\text{P}})}-1)(1-e^{i\eta(\xi_s)})},
\label{Suxyn}
\end{equation} 
and for $\yy\leq0$, 
\begin{equation}
\uu_{\xx,\yy}|_s\sim-\sqrt{\frac{i}{R|{\zphi}_1''(\alpha_s)|}}\frac{C_0}{\sqrt{2\pi}}\mathcal{G}(\varz_s)\frac{e^{i\eta(\xi_s)}e^{iR{\zphi}_1(\xi_s)}}{(1-e^{i(\xi_s-\xi_{\text{P}})})(1-e^{i\eta(\xi_s)})},
\label{Suxyz}
\end{equation}
as $\xi_hR\to\infty$,
where
\begin{eqnarray}
\mathcal{K}(e^{-i\xi})&=&\mathbf{a}^\top(\mathbf{K}_{-}(e^{-i\xi})\mathbf{K}_{-}^{-1}(\xi_{\text{P}})\mathbf{f})
\label{defK}\\
\text{and }
\mathcal{G}(e^{-i\xi})&=&\mathbf{b}^\top(\mathbf{K}_{-}(e^{-i\xi})\mathbf{K}_{-}^{-1}(\xi_{\text{P}})\mathbf{f}).
\label{defG}
\end{eqnarray}
Note that the contribution of the pole $\alpha_{\text{P}}$ is not included in above expressions which accounts for only the saddle point, so that the expressions \eqref{Suxyn} and \eqref{Suxyz} give the approximate diffracted far-fields.

The far-field obtained using the numerical scheme, on a $(2 N_{grid}+1)\times (2 N_{grid}+1)$, is compared with that obtained using semi-analytic. The magnitude of the far-field is plotted in Figure \ref{twocracks} 
for $\NN=4$ and 
for $\MM=0,1,2$.
The semi-analytical results are shown in blue while the numerical results are shown in black. It can be seen that the deviation between the numerical solution and the far-field approximation of the asymptotic Wiener--Hopf factorization based solution is depending on the angle $\theta$. This is mainly due to the presence of asymptotic Wiener--Hopf factors which need certain numerical evaluation of certain contour integral. The anomalous blue dots in all three parts of Figure \ref{twocracks} 
which appear away from a smooth curve is a result of the specific computation of the contour integrals. It has been also observed that as the edges offset $\MM$ increases the agreement of the two results decreases but this is expected as it violates the premise of asymptotic factorization.

\subsection{Low frequency approximation}
In the discrete model adopted here, there are two length scales, the wavelength $2\pi/\kk_1$ (recall \eqref{wavenumber}) of the incoming wave and the square lattice spacing $b$ (see Figure \ref{cracks}). 
As the wavelength $2\pi/\kk_1$ becomes large compared to $b$, the so called continuum limit is obtained, which can also be perceived as a low frequency approximation in the case of assumed square lattice model.
\begin{figure}[htb!]
\centering
\includegraphics[width=\linewidth]{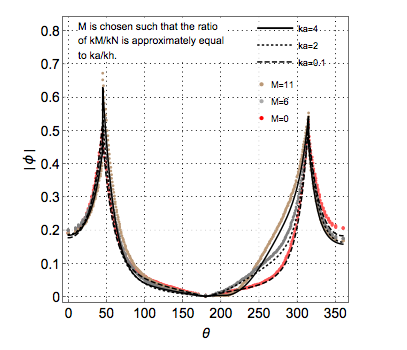}
\caption{Modulus of the diffracted field $|{\zphi}|$ against the observation angle $\obsang$ for angle of incidence $\incang=45^\circ$, $kr=30$, $kh=\pi/2$, $ka=0.1, 2, 4$ (from left to right). The parameters are chosen as given in \cite{abPlatesII}. The thick black curve shows the exact solution obtained by \cite{abPlatesI,abPlatesII} while the grey curve shows the numerical low frequency approximation of the discrete two staggered crack problem. The grey curves are obtained for $\oo=0.35$, $\NN=4$, $\MM=0, 6, 11$ (from left to right), square grid size, $N_{grid}=448$, $N_{pml}=270$. The radius of the discrete circle on the square lattice is taken to be 71 according to the relation \eqref{polarcoord}.}
\label{twocrack_limit_abs}
\end{figure}
The continuum limit of the discrete Helmholtz equation is the continuous Helmholtz equation \cite{Collatz,sConti}. From the perspective of the continuum model, the solutions obtained by \cite{abPlatesI,abPlatesII} can be seen to be approximated via a lattice formulation and the solution of discrete scattering due to the two staggered cracks. 
The numerical solution of the discrete Helmholtz equation for the two staggered cracks on the $(2 N_{grid}+1)\times (2 N_{grid}+1)$ square grid is compared with the exact solution of the continuum model \cite{abPlatesI, abPlatesII} in Figure 
\ref{twocrack_limit_abs}. 
Specifically, the far-field obtained in the two problems has been plotted against the observation angle $\obsang$. The far-field in the case of the discrete problem is obtained as the displacement of the particles on a (large) discrete circle on the square grid $\mathfrak{H}$. In Figure 
\ref{twocrack_limit_abs}, 
the black and grey curves show the far-fields in the known continuum solution \cite{abPlatesI,abPlatesII} and the discrete problem, respectively. It can be seen that for small frequency, the two curves almost coincide.

\section{Conclusion}
\label{conc}
The scattering of a plane time-harmonic wave on a square lattice by two semi-infinite, staggered cracks is considered. The problem is formulated as coupled discrete Wiener-hopf equations. The Wiener--Hopf kernel involved is a $2\times2$ matrix kernel. For multiplicative factorisation of the matrix kernel, an asymptotic method is adapted for a circular contour. The Wiener--Hopf technique is used to solve the coupled Wiener--Hopf equations after obtaining the multiplicative factorisation of the matrix kernel using the asymptotic method. 
The approximate far-fields are obtained using the stationary phase method
and compared with the numerical results. The low frequency approximation of the discrete problem is also shown graphically using the numerical scheme. 
Although the relevant details are omitted, the numerical results for the special case of zero offset, 
using the matrix formulation of this paper have been verified against the exact solution recently presented by \cite{Bls8pair1}.

{\section*{Acknowledgement}
GM acknowledges MHRD (India) and IITK for providing financial assistance in the form of Senior Research Fellowship.
BLS acknowledges the partial support of SERB MATRICS grant MTR/2017/000013.}

\printbibliography

\begin{appendix}
\section{Further simplification and factorization of $G_1$ and $G_2$}
\label{app_elementary}
For illustration of the further manipulations it is assumed that {\em $\NN$ is even}. 
Let 
\begin{equation}
\Nhalf=\NN/2.
\label{Nhalf}
\end{equation}
The function $G_1$ can be written as
\begin{equation}
G_1(\varz)=\lambda^{\Nhalf}(\varz)(\lambda^{\Nhalf}(\varz)+\lambda^{-\Nhalf}(\varz)).
\label{simpleGone}
\end{equation}
Writing $\lambda(\varz)=e^{i\eta(\varz)}$, the second factor, i.e., $\lambda^{\Nhalf}(\varz)+\lambda^{-\Nhalf}(\varz),$ can be written as
$e^{i\eta(\varz)\Nhalf}+e^{-i\eta(\varz)\Nhalf}=
2 \cos\eta(\varz)\Nhalf.$
By the definition of Chebyshev polynomial of the First Kind \cite{mason,sCheby}, $T_n(\ctheta)=\cos{n\eta}$, when $\ctheta=\cos\eta$,
this factor can be modified as
$\lambda^{\Nhalf}(\varz)+\lambda^{-\Nhalf}(\varz)=2T_\Nhalf(\ctheta(\varz)).$
By substituting this expression in \eqref{simpleGone}, the function $G_1$ can modified to the form
\begin{equation}
G_1(\varz)=2\lambda^{\Nhalf}(\varz)T_\Nhalf(\ctheta(\varz)).
\label{SimpleGone}
\end{equation}
Similarly, consider the function $G_2$. This function can be written as
\begin{equation}
G_2(\varz)=-\lambda^{\Nhalf}(\varz)(\lambda^{\Nhalf}(\varz)-\lambda^{-\Nhalf}(\varz))=-2i\lambda^{\Nhalf}\sin{(\eta(\varz)\Nhalf)}.
\label{simplegtwo}
\end{equation}
Using the definition of Chebyshev polynomial of the Second Kind \cite{mason,sCheby}, $U_n(\ctheta)={\sin{(n+1)\eta}}/{\sin\eta}$, when $\ctheta=\cos{\eta}$, the above expression \eqref{simplegtwo} can be written as
\begin{equation}
G_2(\varz)=-2i\lambda^\Nhalf(\varz)\sin\eta(\varz)U_{\Nhalf-1}(\ctheta(\varz))=-2i\lambda^\Nhalf(\varz)(\frac{\lambda(\varz)-\lambda^{-1}(\varz)}{2i})U_{\Nhalf-1}(\ctheta(\varz)).
\label{simgtwo}
\end{equation}
Using the definitions \eqref{lambda} and \eqref{hrlambda}, and the identity, $\lambda^{-1}(\varz)-\lambda(\varz)=r(\varz)h(\varz)$, the expression \eqref{simgtwo} can be rewritten as
\begin{equation}
G_2(\varz)=\lambda^\Nhalf(\varz)r(\varz)h(\varz)U_{\Nhalf-1}(\ctheta(\varz)).
\label{simpleGtwo}
\end{equation}
The polynomials $T_\Nhalf$ and $U_{\Nhalf-1}$ can be written in terms of their zeros \cite{sCheby,handbook}. 
In \eqref{SimpleGone} and \eqref{simpleGtwo}, $\ctheta(\varz)=\frac{\QQ(\varz)}{2}$.
Thus, $T_\Nhalf$ is written in a product form:
$T_\Nhalf(\ctheta(\varz))=2^{\Nhalf-1}\prod\nolimits_{n=1}^{\Nhalf}[\ctheta(\varz)-\cos\frac{(2n-1)\pi}{2\Nhalf}],$
which can be modified to
\begin{equation}
T_\Nhalf(\ctheta(\varz))=2^{\Nhalf-1}\prod\nolimits_{n=1}^{\Nhalf}[\frac{\QQ(\varz)}{2}-\cos\frac{(2n-1)\pi}{2\Nhalf}]
\label{tkprod}
\end{equation}
Let $\zphi_{n-1}=\frac{(2n-1)\pi}{2\Nhalf}$, in \eqref{tkprod}, therefore,
$T_\Nhalf(\ctheta(\varz))=2^{-1}\prod\nolimits_{n=1}^{\Nhalf}[(4-\varz-\varz^{-1}-\oo^2)-2\cos\zphi_{n-1}].$
Using trigonometry, the same expression can be modified and rewritten as
\begin{equation}
\begin{aligned}
T_\Nhalf(\ctheta(\varz))
&=2^{-1}\prod\nolimits_{n=1}^{\Nhalf}[\QQ(\varz)-2+4\sin^2\frac{\zphi_{n-1}}{2}].
\label{TKappa}
\end{aligned}
\end{equation}
Similarly, the function $U_{\Nhalf-1}$ is written in a product form \cite{handbook}:
$U_{\Nhalf-1}(\ctheta(\varz))=2^{\Nhalf-1}\prod\nolimits_{n=1}^{\Nhalf-1}[\ctheta(\varz)-\cos\frac{n\pi}{\Nhalf}].$
Let $\zphi_n=\frac{n\pi}{\Nhalf}$ 
and using the similar manipulations as in case of $T_\Nhalf$, we can write\
\begin{equation}
U_{\Nhalf-1}(\ctheta(\varz))=\prod\nolimits_{n=1}^{\Nhalf-1}[(\QQ(\varz)-2)+4\sin^2\frac{\zphi_n}{2}].
\label{UKappa}
\end{equation}

Let $\mathcal{F}(\varz)=\QQ(\varz)-2+4\sin^2\frac{{\zphi}}{2}$. The zeros of $\mathcal{F}(\varz)$ are $\varz_F({\zphi})$ and $\varz_F^{-1}({\zphi})$, where ($|\varz_F({\zphi})|<1$) (see \cite{sWaveguide})
\begin{equation}
\varz_F({\zphi})=\frac{1}{2}(2+4\sin^2\frac{{\zphi}}{2}-\oo^2\pm\sqrt{(2+4\sin^2\frac{{\zphi}}{2}-\oo^2)^2-4}).
\label{poleszf}
\end{equation}
Thus, the function $\mathcal{F}(\varz)$ is written in terms of its zeroes as
\begin{equation}
\mathcal{F}(\varz;\varz_F)=\varz_F^{-1}(1-\varz_F\varz)(1-\varz_F\varz^{-1}),
\label{functionF}
\end{equation}
and therefore, the factors of $\mathcal{F}$ in the two regions of the complex $\varz$-plane are obtained by writing 
\begin{equation}
\mathcal{F}_{\pm}(\varz;\varz_F)=\varz_F^{-1/2}(1-\varz_F\varz^{\mp1}).
\label{Ffactorsp}
\end{equation}
Then, using the definition \eqref{poleszf}, the product form of $G_1$ and $G_2$, i.e., \eqref{TKappa} and \eqref{UKappa}, respectively, can be modified to 
\begin{equation}
T_\Nhalf(\ctheta(\varz))=2^{-1}\prod\nolimits_{n=1}^{\Nhalf}[\mathcal{F}(\varz;\varz_F(\zphi_{n-1})]; \quad\quad U_{\Nhalf-1}(\ctheta(\varz))=\prod\nolimits_{n=1}^{\Nhalf-1}[\mathcal{F}(\varz;\varz_F(\zphi_n)].
\label{TUkappa}
\end{equation}
Therefore, the functions $G_1$ and $G_2$ can be written in terms of the product form of $T_\Nhalf$ and $U_{\Nhalf-1}$, respectively, that is,
\begin{eqnarray}
G_1(\varz)&=&K(\varz)\prod\nolimits_{n=1}^{\Nhalf}[\mathcal{F}(\varz;\varz_F(\zphi_{n-1})];\\G_2(\varz)&=&K(\varz)r(\varz)h(\varz)\prod\nolimits_{n=1}^{\Nhalf-1}[\mathcal{F}(\varz;\varz_F(\zphi_n)].
\label{GoneGTwo}\\
\text{where }
K(\varz)&=&\lambda^{\Nhalf}(\varz).
\label{defKfac}
\end{eqnarray}
The relation, $\lambda(\varz)=e^{i\eta(z)}$, gives $K(\varz)=e^{i\eta(\varz)\Nhalf}$, where $\eta(z)=\arccos{\frac{\QQ(\varz)}{2}}$. If $f(\varz)=\log K(\varz)=i\Nhalf\eta(\varz)=i\Nhalf\arccos{\frac{\QQ(\varz)}{2}}$, then, using the procedure in \cite{noble} (pp. 21),
\begin{equation}
K_\pm(\varz)=\exp(\frac{1}{2\pi i}\oint_{\mathbb{T}}\frac{f(\alpha)}{\alpha-\varz}\,d\alpha),
\label{fadd}
\end{equation}
which have been found numerically in this paper. The function $J(\varz)=r(\varz)h(\varz)$ can be factorised using \eqref{functionF} with
$\zphi=\pi, 0$ in \eqref{poleszf}; in fact,
\begin{equation}
J_\pm(\varz)=(\varz_r\varz_h)^{-1/4}\sqrt{(1-\varz_r\varz^{\mp1})(1-\varz_h\varz^{\mp1})}
\label{Jrhfact}.\end{equation}
(recall the definitions used in \eqref{annulus}).
Using 
\eqref{TUkappa}, \eqref{fadd}, 
the multiplicative factors of the functions $G_1(\varz)$ and $G_2(\varz)$ are written as
\begin{eqnarray}
G_{1\pm}(\varz)&=&K_{\pm}(\varz)\prod\nolimits_{n=1}^{\Nhalf}\mathcal{F}_{\pm}(\varz;\varz_F(\zphi_{n-1})),
\label{Goneplusminus}\\
\text{and }
G_{2\pm}(\varz)&=&K_{\pm}(\varz)J_{\pm}(\varz)\prod\nolimits_{n=1}^{\Nhalf-1}\mathcal{F}_{\pm}(\varz;\varz_F(\zphi_n)),
\label{Gtwoplusminus}
\end{eqnarray}
respectively. 

\end{appendix}
\end{document}